\documentclass[twoside]{dis09}
\usepackage[latin1]{inputenc}
\usepackage[dvips]{graphicx,epsfig,color}
\usepackage{wrapfig,rotating}
\usepackage{amssymb,amsmath,array}

\usepackage{url}

\newcommand{\pdhejibo}[1]{${#1}$}

\begin{document}
\title{Measurement of $B_c^\pm$ Mass and Lifetime at LHCb}

\author{Jibo HE\\
(on behalf of the LHCb collaboration)
%
%
\vspace{.3cm}\\
%
  Laboratoire de l'Accélérateur Linéaire, Université Paris-Sud 11, CNRS/IN2P3\\
  Université Paris-Sud 11, Bâtiment 200, 91898 Orsay cedex, France, Jibo.He@cern.ch
}

\maketitle

\begin{abstract}
  The $B_c^\pm$ mass and lifetime measurements using the decay
  $B_c^\pm \to J/\psi \pi^\pm$ at the LHCb experiment were studied. About 310 signal
  events are expected for a data set which corresponds to an integrated
  luminosity of 1 fb$^{-1}$, with a $B/S$ ratio equal to 2. Based on these data,
  the $B_c^\pm$ mass and lifetime can be measured with expected statistical
  errors below 2 MeV/$c^2$ and 30 fs, respectively.
\end{abstract}

\section{Introduction}
\label{sec:Introduction}
\nocite{He:dis2009}
The \pdhejibo{B_c^+}\footnote{Charge conjugate states are implied
  throughout this paper, if not specified differently.
}
meson is the ground state of the meson family formed by
two different heavy flavor quarks, the anti-\pdhejibo{b} quark and 
the \pdhejibo{c} quark.
The mass of the \pdhejibo{B_c^+} meson has been estimated theoretically  
to be in a range of 6.2-6.4 GeV/$c^2$~\cite{Brambilla:2004wf}.  
Recently unquenched lattice QCD~\cite{Chiu:2007bc}
gave the most accurate prediction of $M(B_c^+)=6278(6)(4)$ MeV/$c^2$, 
where the first error is statistical, and the second is systematic.
The \pdhejibo{\bar{b}c} mesons are open-flavored and the excited states
below the \pdhejibo{BD} threshold can decay
only through the electromagnetic or hadronic interactions into the ground state,
\pdhejibo{B_c^+}, which can decay only weakly and has a relatively long lifetime.
The estimated \pdhejibo{B_c^+} lifetime is in a range of 0.4-0.8 ps~\cite{Brambilla:2004wf}.
At present, the best prediction is given by QCD 
sum rules~\cite{Kiselev:2000pp}: $\tau(B_c^+)=0.48 \pm 0.05$ ps.
Since the \pdhejibo{\bar{b}c} mesons carry two different heavy flavor quarks,
their production~\cite{Brambilla:2004wf} is more difficult than that of the heavy quarkonia.
Including the contributions from 
the excited states, the cross section of the \pdhejibo{B_c^+} meson at LHC 
was estimated~\cite{Chang:2003cr} to be at a level of 1 $\mu$b, 
and is one order higher than that at Tevatron. This means that O$(10^9)$
\pdhejibo{B_c^+} mesons can be anticipated with 1 fb$^{-1}$ of data at LHC,
which is sufficient to study the \pdhejibo{\bar{b}c} meson family systematically.
 
The \pdhejibo{B_c^+} meson was observed in the semileptonic 
decay modes $B_c^+ \rightarrow J/\psi(\mu^+\mu^-) \ell^+ X\ (\ell=e,\mu)$
by CDF~\cite{Abe:1998wi} at Tevatron.
The measured mass and lifetime are consistent with the theoretical predictions.
More precise measurements of the \pdhejibo{B_c^+} mass have been performed 
recently using the fully reconstructed decay
$B_c^+\rightarrow J/\psi(\mu^+\mu^-) \pi^+$
by CDF~\cite{Aaltonen:2007gv} and D0~\cite{Abazov:2008kv}, giving
$M(B_c^+)=6275.6 \pm 2.9 ({\rm stat.}) \pm 2.5
({\rm syst.})$ MeV/$c^2$ and $M(B_c^+)= 6300 \pm 14 ({\rm stat.}) \pm 5 ({\rm syst.})$MeV/$c^2$ 
respectively.
The \pdhejibo{B_c^+} lifetime measurements using the semileptonic decay 
were also updated with more data,
giving $0.448^{+0.038}_{-0.036} (\rm stat.)\pm 0.032 (\rm sys.)$ ps by
D0~\cite{:2008rba}
and $0.475^{+0.053}_{-0.049} (\rm stat.)\pm 0.018 (\rm sys.)$ ps by
CDF~\cite{CDF:note9294}. 

In this paper, we report our performance study of the \pdhejibo{B_c^+} mass and lifetime
measurements using the exclusive reconstruction of the decay \pdhejibo{B_c^+ \rightarrow
  J/\psi(\mu^+\mu^-) \pi^+} at LHCb.  
The LHCb detector is described elsewhere~\cite{Alves:2008zz}, hereafter
is a summary of the LHCb detector performances which are relevant to this study.
Based on Monte Carlo studies, the relative momentum resolution
$\delta p/p$ was
found to be 0.35\%-0.55\%, depending on the momentum. 
The primary vertex can be measured with a precision of about 10 $\mu$m in
the plane transverse to the beam line, and of about 60 $\mu$m in the beam direction.
The average resolution of the impact parameter respect to the primary vertex
can be parameterized as $\sigma_{\rm IP}=[14 + (35/p_{\rm T})]$ 
$\mu$m where $p_{\rm T}$ is the transverse momentum respect to the
beam direction expressed in GeV/$c$. The muon identification efficiency
was measured to be $\varepsilon(\mu \to \mu)=$94\% with a
misidentification probability $\varepsilon(\pi \to \mu)=$3\%.

\section{Event selection}
\label{sec:EventSelection}
The simulation of the \pdhejibo{pp} collision and 
the subsequent hadronization were performed 
with \textsc{Pythia}~\cite{Sjostrand:2003wg}. 
\pdhejibo{B_c^+} meson 
was generated with a dedicated \pdhejibo{B_c} generator, 
BCVEGPY~\cite{Chang:2006xka}.
The decay of any \pdhejibo{b} hadron produced was done by EvtGen~\cite{Lange:2001uf}.
The response of the detector was simulated by
\textsc{Geant4}~\cite{Agostinelli:2002hh}. 
In this study, the specific \pdhejibo{b\rightarrow J/\psi(\mu^+\mu^-) X} 
(including \pdhejibo{B_d\rightarrow J/\psi X}, \pdhejibo{B_s\rightarrow J/\psi X},
\pdhejibo{B_u\rightarrow J/\psi X}, \pdhejibo{\Lambda_b\rightarrow J/\psi X}),
the inclusive \pdhejibo{J/\psi(\mu^+\mu^-)} and inclusive \pdhejibo{b\bar{b}} MC data samples
were simulated for the background study.

The event selection~\cite{Gao:2008zz} began with selection of
good tracks based on the $\chi^2$ of the track fit,
and requiring a $p_{\rm T}$ greater than 0.5 GeV/$c$.
A pair of two tracks with opposite charges and identified as $\mu$
were used to reconstruct the $J/\psi$ meson. 
\pdhejibo{J/\psi} candidates were required to have  
an invariant mass in the range 
from 3.04 to 3.14 GeV/$c^2$ and a
vertex fit quality of $\chi^2/{\rm ndf}<9$.
Then a track identified as pion from the rest of the event
was combined with the 
\pdhejibo{J/\psi} candidate to reconstruct the \pdhejibo{B_c^+} meson with 
a vertex fit cut of $\chi^2/{\rm ndf}<12$. The impact parameter
significance of the reconstructed \pdhejibo{B_c^+} track
with respect to the production vertex must be less than 3.
For events with more than one primary vertex, the production vertex is
given by the primary vertex to which the \pdhejibo{B_c^+} candidate has the
smallest impact parameter. 
To suppress combinatorial background further, 
$p_{\rm T}(\mu)$ was required to be greater than 1.0 GeV/$c$,
$p_{\rm T}(\pi)$ and $p_{\rm T}(B_c^+)$ were required to be greater
than 1.6 GeV/$c$ and 5.0 GeV/$c$ respectively,
and the impact parameter significance 
of the $\pi$ and $J/\psi$ mesons were required to be 
greater than 3.0 and 3.5 respectively. 

The total reconstruction efficiency for the signal, including the trigger efficiency,
was found to be $(1.01\pm 0.02)$\%. 
In the M(\pdhejibo{B_c^+})$\pm 80$ MeV/$c^2$ (about $\pm 3\sigma$) mass window,
which was regarded as the signal region, the background to signal ratio
$B/S$ with 90\% confidence level is 1 to 2. 
The upper limit of $B/S$ was used in the following study.
One can expect about 310 signal events with 1 fb$^{-1}$ of data
assuming the \pdhejibo{B_c^+} total production cross section
$\sigma(B_c^+)=0.4$ $\mu$b and 
BR(\pdhejibo{B_c^+ \to J/\psi \pi^+})$=1.3 \times 10^{-3}$.

\section{Mass measurement}
To improve the \pdhejibo{B_c^+} mass resolution, the \pdhejibo{J/\psi} mass
constraint vertex fit was used in the mass measurement.
M(\pdhejibo{B_c^+})$\pm 300$ MeV/$c^2$ mass window was used in the
following study. 
The un-binned maximum likelihood method was implemented to extract the
\pdhejibo{B_c^+} mass. The invariant mass distribution of the signal was
modeled by a single Gaussian, 
that of the background was parameterized as a first order polynomial.

\begin{figure}
\centerline{\includegraphics[width=0.5\columnwidth]{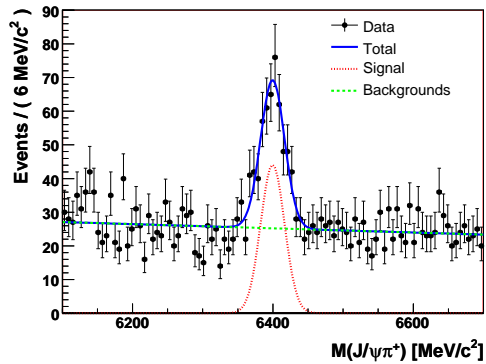}}
\caption{The \pdhejibo{B_c^+} candidate mass distribution 
  together with the result of the fit described in the text. }
\label{Fig:mass}
\end{figure}

For the fit the signal events were taken from the signal data sample 
with the full Monte Carlo simulation and 
the background events were generated by a standalone simulation
using the invariant mass distribution
obtained from the full Monte Carlo simulation.
Both the signal and the background were normalized to 1 fb$^{-1}$ of
data. The fit gives the \pdhejibo{B_c^+} mass to be $M(B_c^{\pm})= 6399.6 \pm 1.7$ (stat.)
MeV/$c^2$, which is in good agreement with the input value of 6400
MeV/$c^2$. The result of the fit is shown in Figure~\ref{Fig:mass}.

\section{Lifetime measurement}
The \pdhejibo{J/\psi} mass constraint vertex fit biases 
the \pdhejibo{B_c^+} decay vertex position
because of the final state radiation in the \pdhejibo{J/\psi} decay
and the energy loss of particles.   
As a consequence, the \pdhejibo{J/\psi} mass constraint vertex fit would bias  
the \pdhejibo{B_c^+} lifetime and was not used in the lifetime study.

A combined mass-lifetime fit was used to measure
the \pdhejibo{B_c^+} lifetime. The invariant mass distributions
for the signal and the background were obtained as in the mass measurement.  
The proper time ($t$) distribution of 
the signal was described by an exponential function smeared by
resolution and multiplied by an acceptance function $\varepsilon(t)$
describing the distortion of $t$ distribution caused by the trigger
and offline event selection. The function form 
was derived from an independent Monte Carlo
sample. The $t$ distribution of the background was parameterized 
by the "KEYS" (Kernel Estimating Your Shapes) 
method~\cite{Cranmer:2000du}. The proper time
resolution ($\sigma_t$) distributions of the signal and the background were also 
considered to avoid the potential bias caused by the so called 
Punzi effects~\cite{Punzi:2004wh}. The fitting framework was tested
with a standalone simulation program.
The pull of the \pdhejibo{B_c^+} lifetime was found to follow a Gaussian distribution 
with $\mu=0.004 \pm 0.010$ and $\sigma=1.0122\pm 0.0072$,
which is consistent with the Normal distribution. 

As shown in Section~\ref{sec:Introduction}, the resolution of the
impact parameter depends on the transverse momentum. 
Therefore, the impact parameter significance and the proper
time acceptance $\varepsilon(t)$ depend on the $p_{\rm T}$(\pdhejibo{B_c^+})
distribution.
In order to evaluate the systematics introduced by this, 
a fit to \pdhejibo{B_c^+}, which was generated with the $p_{\rm T}$
distribution predicted for \pdhejibo{B^+} by \textsc{Pythia}, was made using 
the acceptance function obtained from the \pdhejibo{B_c^+} data generated by BCVEGPY. 
Note that the $p_{\rm T}$(\pdhejibo{B_c^+}) spectrum predicted by BCVEGPY is
harder than $p_{\rm T}$(\pdhejibo{B^+}) spectrum predicted by \textsc{Pythia}.
A standalone simulation study shows that in this case the \pdhejibo{B_c^+} lifetime would be
biased by about 0.023 ps.

To reduce the dependence of the lifetime measurement on the $p_{\rm
  T}$(\pdhejibo{B_c^+}) distribution, the $p_{\rm T}$(\pdhejibo{B_c^+}) was divided
into two intervals: 5-12 GeV/$c$ and $> 12$ GeV/$c$. 
The event selection was re-optimized and for the 
data set with $p_{\rm T}$(\pdhejibo{B_c^+}$)> 12$ GeV/$c$, 
the impact parameter significance
cuts are changed to $> 2.0$, $> 2.5$ and $< 4.0$ for pions,
\pdhejibo{J/\psi} and \pdhejibo{B_c^+} respectively.  
The event selection results of the two $p_{\rm T}$ intervals are summarized
in Table~\ref{tab:SelSumTwoPT}.
The combined mass-lifetime fit applied simultaneously 
for the two $p_{\rm T}$(\pdhejibo{B_c^+}) intervals reduces 
the bias in the lifetime measurement to about 0.004 ps.

\begin{table}
  \centering
  \begin{tabular}{|l| c| c|}
        \hline
        $p_{\rm T}$ intervals of \pdhejibo{B_c^+}                 &  5-12 GeV/$c$  & $\geq$ 12 GeV/$c$ \\\hline
        Total efficiency ~$\varepsilon_{\rm
          tot}$
        (including trigger)     & $(0.34\pm 0.01)$ \%  &
        $(0.86\pm 0.02)$ \% \\\hline
        Signal yield \small{(1 fb$^{-1}$ @ 14 TeV)}                              & $100$ & $260$  \\\hline
        $B/S$ @ 90\% CL                      & 3 to 6           &  0.6
        to 1.2 \\
        \hline
  \end{tabular}
  \caption{Summary of the event selection results in the two $p_{\rm
      T}$(\pdhejibo{B_c^+}) intervals.
    The total efficiency $\varepsilon_{\rm tot}$ in the two $p_{\rm T}$ intervals
    are both defined with respect to the total number of the
    events in the full $p_{\rm T}$ range.}
  \label{tab:SelSumTwoPT}
\end{table}

\begin{figure}[!ht]
  \centering
  \resizebox{1.0\textwidth}{!}{%
    \includegraphics{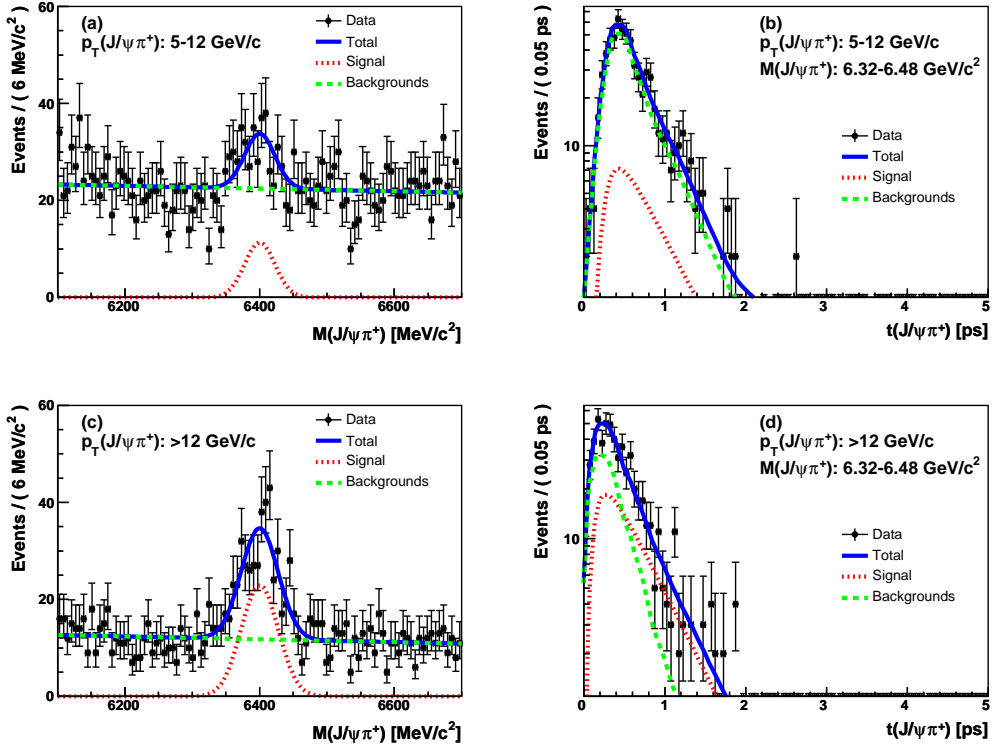}}
    \caption{The results of the fit in the two $p_{\rm T}$(\pdhejibo{B_c^+})
      intervals. The left column shows the invariant mass
      distributions, the right column shows the proper time
      distributions in the signal region.}
    \label{fig:FitResults}
\end{figure}

The fit to the 1 fb$^{-1}$ of data gives \pdhejibo{B_c^+} lifetime to be 
$\tau$(\pdhejibo{B_c^+}) = $0.438 \pm 0.027$ (stat.) ps,
which is in good agreement with the input value of 0.46 ps. 
The results of the fit in the two $p_{\rm T}$(\pdhejibo{B_c^+}) intervals are shown 
in Figure~\ref{fig:FitResults}.

\section{Summary}
In summary, the event selection of the decay \pdhejibo{B_c^+ \to
  J/\psi(\mu^+\mu^-) \pi^+} at LHCb was studied, about 310 signal events are
expected from 1 fb$^{-1}$ of data, with a $B/S$ of 2. 
Based on the selected data sample, the \pdhejibo{B_c^+} mass can be measured
with an expected
statistical error to be below 2 MeV$/c^2$. For the \pdhejibo{B_c^+} lifetime
measurement, to reduce a potential bias caused by the limited
knowledge of the $p_{\rm T}$(\pdhejibo{B_c^+}) distribution, 
the data sample was divided into two $p_{\rm T}$(\pdhejibo{B_c^+}) intervals.
The combined mass-lifetime fit performed in these
two $p_{\rm T}$ intervals simultaneously gives a statistical uncertainty
of the \pdhejibo{B_c^+} lifetime to be below 30 fs.


\begin{footnotesize}

\bibliographystyle{unsrt}
\bibliography{he_jibo}
\end{footnotesize}


\end{document}